\documentstyle[epsf]{l-aa}

\begin{document}

   \thesaurus{04           % A&A Section 04
              (08.12.2;
               10.19.1;
               10.11.1)}
   \title{The contribution of brown dwarfs to the local mass budget of the
          Galaxy}

   \author{B. Fuchs$^{\rm 1}$, H. Jahrei{\ss}$^{\rm 1}$,
           and C. Flynn$^{\rm 2}$}

   \offprints{B. Fuchs, (fuchs@ari.uni-heidelberg.de)}

   \institute{$^{\rm 1}$Astronomisches Rechen-Institut Heidelberg,
              M\"onchhofstr. 12-14, 69120 Heidelberg, Germany\\
              $^{\rm 2}$Tuorla Observatory, V\"ais\"al\"antie 20,
              Piikki\"o, Finland}

   \date{Received 2 June 1998/Accepted  }

   \maketitle
   \markboth{B. Fuchs et al.~: Contribution of brown dwarfs to the
            local mass budget}{B. Fuchs et al.~: Contribution of
            brown dwarfs to the local mass budget}

   \begin{abstract}
   Based on the recent discoveries of free floating brown dwarfs we derive
   estimates of the local mass density of this population of objects.
   Mass density
   estimates from various surveys span the range 0.03 to 0.005 $M_\odot$
   pc$^{\rm -3}$. These estimates are compared with the local mass densities
   of the
   other constituents of the galactic disk and, in particular, with the
   dynamically determined total local mass density. We argue that brown dwarfs
   might indeed contribute significantly to the local mass budget, but that a
   local mass density as high as 0.03 $M_\odot$ pc$^{\rm -3}$ as suggested by
   Ruiz et al.~(1997) is rather unlikely.

   \end{abstract}

%
%________________________________________________________________

\section{Introduction}
The recent discoveries of free floating brown dwarfs by the Calan-ESO proper
motion survey (Ruiz et al.~1997), the DENIS mini survey (Delfosse
et al.~1997, Tinney et al.~1997), the BRI survey (Irwin et al.~1991,
Tinney 1998) and the UK Schmidt deep photographic
survey (Hawkins et al.~1998) have now firmly established the long suspected
existence of a population of single brown dwarfs in the disk of the
Galaxy. Among the many interesting questions raised by these important
discoveries is the local mass density of such objects. Although based at
present on very low number statistics Ruiz et al.~(1997) have
pointed out that brown dwarfs may contribute significantly to the local mass
budget. It is the aim of this note to assess the various density estimates and
relate them to the total local mass density.

\section{Estimates of the local mass density of brown dwarfs}

\subsection{Calan-ESO proper motion survey}

The Calan-ESO proper motion survey has at present led to the discovery
of one brown dwarf, Kelu-1 (Ruiz et al.~1997). The authors estimate
that their effective search volume had a size of 23 pc$^{\rm 3}$, which implies
a number density of 0.04 pc$^{\rm -3}$ or
a mass density of 0.0028 $M_\odot$ pc$^{\rm -3}$, if a typical
mass of a brown dwarf of 0.065 $M_\odot$ is assumed. However, these numbers may
underestimate the actual density. First, the survey detects only stars with
proper motions exceeding 0.25 arcsec/yr, so that some stars with low space
velocities might not be found. Furthermore, the plate limits imply that even
among the very nearby brown dwarfs such as Kelu-1 only the bright
brown dwarfs will be detected. Ruiz et al.~(1997) and Tinney (1998)
argue that these objects have ages less than one Gyr. In order to correct for
both selection effects we assume that the distribution function of brown dwarfs
in phase space
is the same as that of the luminous stars. The velocity distribution is then
described by Schwarzschild distributions, i.e.~three-dimensional Gaussians
with anisotropic axial ratios. As is well known from nearby stars (Wielen
1977), the velocity dispersions of the Schwarzschild distributions increase
with the ages of the stars. We assume the same behaviour for the brown dwarfs,
because the dynamical evolution of a population of brown dwarfs is expected to
be the same as for luminous stars. The velocity distribution of young brown
dwarfs with ages less than 1 Gyr is then modelled as
\begin{eqnarray}
\lefteqn{f({\bf v}) =
\nu_{\rm 0} \int_{\rm 0}^{\rm 1 Gyr}d\tau\frac{r_{\rm sfr}(\tau)}
{\sigma_{\rm U}(\tau) \sigma_{\rm V}(\tau) \sigma_{\rm W}^2(\tau)} }\\
 & & \exp{-\frac{1}{2} \left[
\left(\frac{U-U_\odot}{\sigma_{\rm U}(\tau)}\right)^2 +
\left(\frac{V-V_\odot}{\sigma_{\rm V}(\tau)}\right)^2 +
\left(\frac{W-W_\odot}{\sigma_{\rm W}(\tau)}\right)^2 \right]}\,.
\nonumber
\end{eqnarray}
In Eq.~(1) $U, V, W$ denote the space velocity components of the brown
dwarfs in the direction of the galactic center, the direction of galactic
rotation and towards the galactic north pole, respectively. $\sigma_{\rm U},
\sigma_{\rm V}, \sigma_{\rm W}$ are the corresponding velocity dispersions.
They are assumed to increase with age as
\begin{equation}
\sigma^{2}_{\rm U,V,W}(\tau) = \sigma^{2}_{\rm U0,V0,W0} + C_{\rm U,V,W}
\tau \,,
\end{equation}
like the velocity dispersions of the luminous stars (Wielen 1977). The axial
ratios of the velocity ellipsoid have been chosen according to the most recent
discussion of the kinematics of nearby stars by Jahrei{\ss} \& Wielen (1997) as
$\sigma_{\rm U}^2:\sigma_{\rm V}^2:\sigma_{\rm W}^2=2:1:0.57$. The
diffusion coefficients $C_{\rm U,V,W}$ follow the same ratios and have an
absolute value of
 $C_{\rm U}+C_{\rm U}+C_{\rm W}$ = 6 $\cdot$ 10$^{-7}$ (km/s)$^2$/yr.
Initial velocity dispersions of
$\sigma^{2}_{\rm U0} + \sigma^{2}_{\rm V0} + \sigma^{2}_{\rm W0}$ = 100
(km/s)$^2$ have been assumed
(Wielen 1977, Jahrei{\ss} \& Wielen 1997). For the solar motion with
respect to the local standard of rest standard values of ($U_\odot, V_\odot,
W_\odot$) = (9, 12, 7) km/s have been adopted. The velocity distribution $f$ is
calculated by a superposition of Schwarzschild distributions for each
generation of stars weighted by the star formation rate $r_{\rm sfr}$, defined
per surface density, which we have assumed to be constant with
respect to time. Stars with higher vertical velocity dispersions settle with
larger vertical scale heights in the galactic gravitational potential than
stars with low vertical velocities , so that there will be a thinning-out
of the density of such stars at the galactic midplane. This is modelled in
Eq.~(1) by an extra $\sigma_{\rm W}(\tau)^{-1}$ term. $\nu_{\rm 0}$ takes
care of all the normalization constants.

The expected number density of brown dwarfs in a cone towards the direction of
Kelu-1 ($l$=307.7$^\circ$, $b$=36.8$^\circ$) is given by
\begin{equation}
\nu_{\rm exp} = \left(\frac{1}{3} {\rm d}_{\rm max}\right)^{-3} \int_{\rm
0}^{\rm d_{max}} dr r^2 \int d^3v f({\bf v}),
\end{equation}
where d$_{\rm max}$ is the radial length of the probing volume investigated by
Ruiz et al.~(1997). The observed number density of brown dwarfs in the
cone, $\nu_{\rm obs}$, can be expressed similarly to Eq.~(3), if the
proper motion limit of $\mu \geq$ 0.25 arcsec/yr of the Calan-ESO poper
motion survey is taken into account in the integration over the velocity
components. To implement this proper motion limit one has to transform the
cartesian velocity components $U, V$, and $W$ in Eq.~(3) to radial and
tangential velocities and from that to proper motions. The integrand in
Eq.~(3) depends
then explicitly on the radial distance $r$. We set the upper boundary of the
$r$-integration at d$_{\rm max}$ = 10 pc. This is based on the observation
that Kelu-1 is very similar to DENIS-P\,J1228.2-1547 (Tinney et al.~1997),
implying an absolute magnitude of M$_{\rm R}$ = 19.5 mag,
which is about the plate limit of the Calan-ESO survey (Ruiz et al.~1997).
\footnote{We note that this places Kelu-1 at a distance of 9 pc,
which is less than the `astrometric' distance derived by Ruiz et al.~(1997).
However, we do not follow the authors in assuming that Kelu-1 is
exactly at rest at the local standard of rest, so that its proper motion
solely reflects the solar motion. Indeed, LTT\,5054 and LTT\,5122, which lie in
the same direction as Kelu-1 and have very similar proper motions, but for
which
parallaxes are known, can be shown to be typical disk stars with non-zero
space velocities.} Integrating Eq.~(3) and its modification numerically we
find a ratio of
\begin{equation}
\frac{\nu_{\rm obs}}{\nu_{\rm exp}} = 0.39 \,,
\end{equation}
which changes to 0.28 or 0.44, if d$_{\rm max}$ is increased or decreased by a
factor of 1.6 according to an estimated uncertainty of the distance modulus of
$\pm$1
mag, respectively. Up to now we have considered only bright and young brown
dwarfs. Very likely there are also fainter and older brown dwarfs, which will
contribute to the mass budget as well, although they have not yet been found in
the surveys. Assuming again a constant star formation rate and an age of the
galactic disk of 10 Gyrs we estimate from Eq.~(1)
\begin{equation}
\frac{\nu (\tau \leq 10 {\rm Gyr})}{\nu (\tau \leq 1 {\rm Gyr})} =
\frac{\sqrt{\sigma_{\rm W0}^2 + C_{\rm W}\cdot 10 {\rm Gyr}} - \sigma_{\rm W0}}
{\sqrt{\sigma_{\rm W0}^2 + C_{\rm W}\cdot 1 {\rm Gyr}} - \sigma_{\rm W0}} = 4.1 \,.
\end{equation}
Combining this with the estimate (4) we conclude that the contribution of
brown dwarfs to the local mass density is 10 times the density of brown dwarfs
deduced
straightforward from the Calan-ESO survey. The same correction factor is
given by Ruiz et al.~(1997), although the authors do not explain in
detail how they arrive at that estimate.

\subsection{DENIS mini survey}

The DENIS mini survey with a field size of 230 deg$^2$ has led to the discovery
of three brown dwarf candidates (Delfosse et al.~1977). High resolution
spectroscopy by Tinney et al.~(1997) has clearly confirmed the
brown dwarf nature of DENIS-P\,J1228.2-1547 by detecting a Li absorption
line. However, Tinney et al.~(1997) conclude that the other two
objects must be also very cool low-mass objects. If we adopt the absolute K
magnitude of DENIS-P\,J1228.2-1547 and the plate limit in the K band given by
Delfosse et al.~(1997), we find that the search volume of the DENIS mini survey
for brown dwarfs is 162 pc$^3$. This implies a local number density of brown
dwarfs of
0.019 pc$^{-3}$. In order to account for the fainter and older brown dwarfs,
which have not yet been detected by the DENIS survey, the correction factor
given in Eq.~(5) has to be applied.

\subsection{BRI survey}

High resolution spectroscopy of very late type stars of the BRI survey (Irwin
et al.~1991) by Tinney (1998) has led to the identification of the
brown dwarf LP\,944-20 = BRI\,0337-3535. Tinney (1996, 1998) has also measured
the parallax, proper motion and radial velocity of this brown dwarf. Its age is
estimated as about 500 Myrs (Tinney 1998). The low space velocity of 7 km/s is
consistent with such a young age (Wielen 1977). The BRI survey covers 1000
deg$^2$ and its plate limit is at 19.0 mag in the R band. In the NLTT catalogue
(Luyten 1979) the apparent R magnitude of LP\,944-20 is given as R = 17.5
mag. This implies that the search volume of the BRI survey for brown dwarfs is
99 pc$^3$. The local number density of brown dwarfs is then according to this
determination 0.01 pc$^{-3}$. Since LP\,944-20 is younger than the brown dwarfs
found in the other surveys (Tinney 1998), the correction factor for taking into
account the fainter and older brown dwarfs is 6.8 (cf.~Eq.~(5)).

\subsection{UK Schmidt survey}

Spectroscopy of very red stars of the UK Schmidt survey of ESO/SERC field 287
(Hawkins et al.~1998) has led to the identification of three brown dwarf
candidates at distances
between 37 and 48 pc. The plate limit is at 23.1 mag in the R band, and the
field area is 25 deg$^2$, although crowding reduces the effective area
of the survey by as much as a factor of 4. Using the apparent R magnitudes and
the parallaxes given by Hawkins et al.~(1998) we estimate that the survey is
complete up to a distance of 50 pc. This implies an effective search volume
of 80 pc$^3$ and the local number density of brown dwarfs is then
0.038 pc$^{-3}$. Hawkins et al.~(1998) assume that the ages of the brown
dwarfs, which they have found, are about one Gyr. Thus the correction
factor for the older brown dwarfs given in Eq.~(5) has to be applied.

\section{Discussion and Conclusions}

The four surveys have led to number density estimates of brown dwarfs
of 0.46 pc$^{-3}$, 0.076 pc$^{-3}$, 0.069 pc$^{-3}$, and 0.15 pc$^{-3}$
respectively. Assuming again a typical mass of a brown dwarf of 0.065 $M_\odot$
this corresponds to mass densities of 0.03 $M_\odot$ pc$^{-3}$, 0.0049
$M_\odot$ pc$^{-3}$, 0.0045 $M_\odot$ pc$^{-3}$ and 0.01 $M_\odot$ pc$^{-3}$,
respectively. We note, however, that despite the large differences the
estimates are still statistically consistent within two standard
deviations. This can be shown by integrating the Poisson probability function
with respect to the mean value at a given number of N realizations.  The range
of mean values (x$_{\rm l}$, x$_{\rm u}$) of Poisson distributions from which
the N realizations have been drawn with a probability of 95\%, for instance, is
then given by
\begin{equation}
\int_{\rm 0, x_u}^{\rm x_l,\infty} \frac{x^{\rm N}}{N!} e^{-x} dx = 0.025 \,.
\end{equation}
If N=1, (x$_{\rm l}$, x$_{\rm u}$) = (0.24, 5.6) and in the case of N=3
(x$_{\rm l}$, x$_{\rm u}$) = (1.1, 8.8). This shows immediately that the
statistical uncertainties of the density estimates of brown dwarfs overlap at
two standard deviations.

The high estimate of the space density of Kelu-1 type objects, leads us to ask
if they would be detectable in deep star count data taken with the Space
Telescope (HST).  We first consider star counts in the Hubble Deep Field (HDF)
(Flynn et al.~1996).  No very red stars, i.e.~stars which appeared only in the
I-band images and not in the V-band images, were detected in the HDF to a
limiting magnitude of I=26.3. The number density implied by Kelu-1 itself is
0.1 pc$^{-3}$, while figuring in the older brown dwarfs associated with Kelu-1
raises the estimated number density to 0.4 pc$^{-3}$. Taking
the absolute magnitude of Kelu-1 like other brown dwarfs as M$_I=16.6$
and their local density to be 0.4 pc$^{-3}$ and assuming
an exponential scale height of 300 pc, we estimate that 6 Kelu-1 type objects
would have appeared in the 4.4 square arcminute HDF, which is marginally
inconsistent with none being observed, while only 0.5 would be expected if they
have an exponential scale height of 100 pc, which is consistent with none
observed. Stronger limits can be obtained using faint HST star counts in the
Groth Strip (Gould et al.~1997), which covers 114.0 square arcminutes to a
limiting I-band magnitude of $I=23.8$. In this field we would expect 5 Kelu-1
type objects for a scale height of 100 pc and 15 for a scale height of 300 pc,
whereas only 1 very red object (V-I$>$5.0) was detected. However, this assumes
that all of the 0.4 pc$^{-3}$ local space density of Kelu-type brown dwarfs
would be as bright
as Kelu-1 itself, whereas most will be older and fainter. For example, in a
two-component toy model, containing a young component with a space density of
0.1 pc$^{-3}$, scale height 100 pc and M$_I=16.6$ and an old component with a
space  density of 0.3 pc$^{-3}$, scale height 300 pc and M$_I=18.7$ we would
expect
circa 1 star from each component in the Groth strip, consistent with 1 observed
red star. We conclude that faint star counts in HDF and the Groth strip are
unable to rule out the high space density of Kelu-1 type objects measured by
Ruiz et al.~(1997).

These density estimates can be compared with measurements of the densities of
the other constituents of the galactic disk. Jahrei{\ss} \& Wielen (1997)
derive from a discussion of the most recent data of nearby stars that the
mass density of luminous stars in the solar neighbourhood is 0.039 $M_\odot$
pc$^{-3}$. Thus, if the mass density estimate of brown dwarfs based on the
Calan-ESO survey is correct, brown dwarfs contribute almost as much to the
local mass budget as the luminous stars.  The mass density estimates of brown
dwarfs
can be also compared with dynamically determined local mass densities. Such
measurements include all gravitating matter. Fuchs \& Wielen (1993) and Flynn
\& Fuchs (1994) have used samples of K dwarfs and K giants, respectively, to
measure the local slope of the K$_{\rm z}$ force law. Both measurements have
led to a total local mass densities of 0.1 $M_\odot$ pc$^{-3}$ with an
estimated
uncertainty of 20\%.A repetition of the measurement using improved data of K
giants indicates a total local mass density at the lower end of this range
(Flynn
\& Fuchs, in preparation). Cr\'{e}z\'{e} et al.~(1997) have used recently
Hipparcos observations of A stars and determine a value of the total local mass
density of 0.076 $\pm$ 0.015 $M_\odot$ pc$^{-3}$. The local surface density of
interstellar gas in the form of cold HI and H$_{\rm 2}$ is 6 $M_\odot$
pc$^{-2}$ (Dame 1993). If corrected for the presence of heavier elements and
folded with a vertical scale height of 100 pc, this implies a local mass
density of interstellar gas of 0.037 $M_\odot$ pc$^{-3}$. The mass density of
warm HI and ionized interstellar gas is more difficult to assess, but is
probably only 0.003 $M_\odot$ pc$^{-3}$ (Kuijken \& Gilmore 1984). We conclude
from this discussion that there is evidence for a local mass density of brown
dwarfs of the order of
0.01 $M_\odot$ pc$^{-3}$, which would seem to fill the gap between dynamical
mass determinations and the mass density of so far identified stellar and
interstellar matter,
whereas the mass density estimate by Ruiz et al.~(1997)
(1997) seems to be on the high side. Hopefully, further discoveries of brown
dwarfs, such as announced by the 2MASS survey (Kirkpatrick et al.~1998), will
clarify this issue.

%
%______________________________________________________________
\begin{acknowledgements}
We gratefully acknowledge valuable discusions with E.~Kerins and R.~Wielen.
\end{acknowledgements}


\begin{thebibliography}{}

\bibitem{} Cr\'{e}z\'{e} M., Chereul E., Bienaym\'{e} O., Pichon C., 1997,
A\&A 329, 920

\bibitem{} Dame T.M., 1993, in: Back to the Galaxy, eds.\ S.S. Holt, F. Verter,
AIP Conf. Proc. 278, 267

\bibitem{} Delfosse X., Tinney C.G., Forveille T. et al., 1997, A\&A 327, L25

\bibitem{} Flynn C., Fuchs B., 1994, MNRAS, 270, 471

\bibitem{} Flynn, C, Gould, A, Bahcall, J, 1996, ApJ, 466, L55

\bibitem{} Fuchs B., Wielen R., 1993, in: Back to the Galaxy, eds.\ S.S. Holt,
F. Verter, AIP Conf. Proc. 278, 580

\bibitem{} Gould, A, Bahcall, J, Flynn C., 1997, ApJ, 482, 913

\bibitem{} Hawkins, R.S., Ducourant, C, Jones, H.R.A., Rapaport, M. 1998, MNRAS
294, 505

\bibitem{} Irwin M., McMahon R.G., Hazard C., 1991, in: The space distribution
of quasars, ed.\ D. Crampton, Astron. Soc. Pac. Conf. Ser. 21, 117

\bibitem{} Jahrei{\ss} H., Wielen R., 1997, in: HIPPARCOS Venice '97, eds.\
M.A.C. Perryman, P.L. Bernacca, ESA SP-402, 675

\bibitem{} Kirkpatrick J.D., Cutri R.M., Nelson B. et al., 1998, BAAS 30 No.~
2, Abstr. 54.04

\bibitem{} Kuijken K., Gilmore G., 1984, MNRAS 239, 605

\bibitem{} Luyten W.J., 1979, NLTT Catalogue, Univ. of Minnesota, Minneapolis

\bibitem{} Ruiz, M.T., Leggett, S.K., Allard F., 1997, ApJ 491, L107

\bibitem{} Tinney C.G., 1996, MNRAS 281, 644

\bibitem{} Tinney C.G., 1998, MNRAS, in press, astroph/9801171

\bibitem{} Tinney C.G., Delfosse X., Forveille T., 1997, ApJ 490, L95

\bibitem{} Wielen R., 1977, A\&A 60, 263

\end{thebibliography}
\end{document}